\documentclass[pra,aps,twocolumn,reprint,showpacs,superscriptaddress,floatfix]{revtex4-1}
\usepackage[utf8]{inputenc} 
\usepackage{graphicx}
\usepackage{amsmath}
\usepackage{rotating}
\usepackage{units}
\newcommand{\be}{\begin{equation}}
\newcommand{\ee}{\end{equation}}
\newcommand{\bea}{\begin{eqnarray}}
\newcommand{\eea}{\end{eqnarray}}

\usepackage{color}

\begin{document}
\title{$s$-wave scattering lengths of the strongly dipolar bosons $^{162}$Dy and $^{164}$Dy}
\author{Yijun Tang}
\affiliation{Department of Physics, Stanford University, Stanford CA 94305}
\affiliation{E.~L.~Ginzton Laboratory, Stanford University, Stanford CA 94305}
\author{Andrew Sykes}
\affiliation{JILA, University of Colorado and National Institute of Standards and Technology, Boulder CO 80309}
\author{Nathaniel Q. Burdick}
\affiliation{E.~L.~Ginzton Laboratory, Stanford University, Stanford CA 94305}
\affiliation{Department of Applied Physics, Stanford University, Stanford CA 94305}
\author{John L. Bohn}
\affiliation{JILA, University of Colorado and National Institute of Standards and Technology, Boulder CO 80309}
\author{Benjamin L. Lev}
\affiliation{Department of Physics, Stanford University, Stanford CA 94305}
\affiliation{E.~L.~Ginzton Laboratory, Stanford University, Stanford CA 94305}
\affiliation{Department of Applied Physics, Stanford University, Stanford CA 94305}

\date{\today}

\begin{abstract}
We report the measurement of the deca-heptuplet $s$-partial-wave scattering length $a$ of two bosonic isotopes of the highly magnetic element, dysprosium:  $a=122(10)a_0$ for $^{162}$Dy and $a=92(8)a_0$ for $^{164}$Dy, where $a_0$ is the Bohr radius.  The scattering lengths are determined by the cross-dimensional relaxation of ultracold gases of these Dy isotopes at temperatures above quantum degeneracy.   In this temperature regime, the measured rethermalization dynamics can be compared to simulations of the Boltzmann equation  using a direct-simulation Monte Carlo (DSMC) method employing the anisotropic differential scattering cross section of dipolar particles.

\end{abstract}

 \maketitle
 
\section{Introduction}

In the study of ultracold atomic collisions, the scattering length $a$ is a simple parameter that characterizes the contact-like pseudo-potential approximation of the van der Waals potential~\cite{pethick2002bose}\footnote{In dipolar gases, this scattering length selects the isotropic part of the scattering amplitude and depends on the dipole moment~\cite{Ronen06,YiYou01,*YiYou00}:  Therefore, the scattering lengths measured here are Dy's pseudo-potential scattering lengths $a(\mu)$ at the strength of Dy's dipole moment $\mu$.  For fixed dipole moment $\mu$, we may define $a \equiv a(\mu)$, and the renormalization correction $a(\mu)/a(0)$ has been calculated to be only on the order of a few percent for Dy's magnitude of $\mu$~\cite{Ronen06}.}.
By abstracting away microscopic details, this number encapsulates the essential physics needed to predict the cross section of atoms whose collision channel is dominated by an $s$ partial wave.  Knowledge of $a$ allows one to predict the mean-field energy of a Bose-Einstein condensate (BEC). Manipulating $a$ via a Fano-Feshbach resonance  provides interaction control~\cite{Chin2010}, which can increase evaporation efficiency for BEC production~\cite{Cornish:2000,Weber:2003,Tang:2015} or provide access to strongly interacting gases and gases that emulate interesting many-body Hamiltonians~\cite{Bloch:2008}.

Given the importance of the $s$-wave scattering length, it is desirable to know its value for the highly magnetic and heavy open-shell lanthanide atom dysprosium (Dy), whose three high-abundance bosonic isotopes have recently been Bose-condensed~\cite{Lu2011,Tang:2015}.  However, Dy has a highly complex electronic structure: an open $f$-shell submerged beneath closed outer $s$-shells.  The four unpaired $f$ electrons give rise to a total electronic angular momentum $J=L+S=8$, with an orbital angular momentum $L=6$ and electronic spin $S=2$. (Bosonic Dy has no nuclear spin $I=0$ and hence has no hyperfine structure.)  The complexity of Dy's electronic structure---possessing 153 Born-Oppenheimer molecular potentials, electrostatic anisotropy, and a large dipole moment ($\mu = 9.9326952(80)$ Bohr~magnetons~\cite{Martin:1978})---renders  calculating collisional parameters  challenging~\cite{Kotochigova2011}.  Therefore, as with all but the lightest atoms,  determination of the scattering length must rely on  experimental measurements~\cite{pethick2002bose}. 

One well-known technique often used to probe the collisional properties of ultracold atoms is the cross-dimensional relaxation method~\cite{Monroe:1993}. Such experiments usually begin with a cloud of atoms in thermal equilibrium. Then extra energy is suddenly added to the cloud along one of the trap axes to create an energy imbalance. This may be accomplished by diabatically increasing the trap frequency in that direction. One can then extract the elastic cross section of the colliding particles by measuring the rate at which this energy redistributes among all three trap axes. 

For bosonic alkali atoms, whose collision interaction is dominated by $s$-wave scattering at ultracold temperatures, i.e.,  below the $d$-wave centrifugal energy barrier, the elastic cross section is directly related to the scattering length~\cite{pethick2002bose}. However, the scattering in ultracold bosonic Dy gases is strongly affected by the magnetic dipole-dipole interaction (DDI). In contrast to the short-ranged, isotropic van der Waals interaction, the DDI is long-ranged and highly anisotropic:
\begin{equation}
U_{\rm{dd}}({\bf r})=\frac{\mu_0\mu^2}{4\pi}\frac{1-3\cos^2{\theta}}{|{\bf r}|^3},
\end{equation}
where $\mu_0$ is the vacuum permeability, {\bf r} is the relative position of the dipoles, and $\theta$ is the angle between $\bf r$ and the dipole polarization direction. Scattering due to the DDI has been calculated to be universal in the ultracold regime, meaning that it does not depend on the microscopic details of the colliding particles~\cite{Bohn2009}. Such scattering can be characterized by a single parameter, the dipole length scale
\begin{equation}
a_d=\frac{\mu_{0} \mu^2m}{8\pi\hbar^2},
\end{equation}
where $m$ is the single-particle mass~\cite{Bohn2009,DipoleLength:footnote}. The universal nature of the DDI has  been observed  for both elastic~\cite{Hensler2003,Pasquiou:2010ii,Lu2012,Aikawa2013} and inelastic collisions~\cite{Burdick:2015}.  The remaining non-universal part of scattering resides in the scattering length, whose value varies from atom to atom.

The goal of this work is to measure $a$, which includes the small dipolar contribution [2], by accounting for the DDI in the total Dy-Dy elastic cross section. This is achieved by comparing the measured cross-dimensional relaxation of an ultracold gas of Dy to numerical simulations in which the DDI's contribution to the cross section is well understood~\cite{BohnJin_PRA_89_022702_2014}.  The simulation of the nonequilibrium dynamics of ultracold dipolar gases in realistic experimental situations is made possible by a recently developed direct-simulation Monte Carlo (DSMC) method that solves the Boltzmann equation with the full dipolar differential scattering cross section~\cite{SykesBohn_PRA_91_013625_2015}. This numerical method has proven successful in describing the rethermalization of a cloud of fermionic erbium atoms driven out of equilibrium~\cite{Aikawa:2014}. Here we apply these simulation tools to  bosonic $^{162}$Dy and $^{164}$Dy undergoing cross-dimensional relaxation and extract the deca-heptuplet $s$-wave scattering length $a$ for both isotopes in their maximally stretched ground state $|J=8,m_J=-8\rangle$.

\section{The cross-dimensional relaxation experiment}

Preparation of ultracold Dy gases is discussed in a previous work~\cite{Tang:2015}. Dysprosium atoms in an atomic beam generated by a high-temperature effusive cell are loaded into a magneto-optical trap (MOT) via a Zeeman slower, both operating at 421~nm. For further cooling, the atoms are loaded into a blue-detuned, narrow-linewidth MOT at 741~nm. We typically achieve trap populations of $4\times10^7$ $^{162}$Dy or $^{164}$Dy atoms at $T\approx2$~$\mu$K.   The atoms confined within this narrow-line, blue-detuned MOT are spin-polarized in $|J=8,m_J=+8\rangle$. They are subsequently loaded into a single-beam 1064-nm optical dipole trap (ODT). Once in the ODT, the atoms are transferred to the absolute electronic ground state $|J=8,m_J=-8\rangle$ by radio-frequency-induced adiabatic rapid passage. We then perform forced evaporative cooling in two differently optimized crossed optical dipole traps (cODT) formed by three 1064-nm beams. The first cODT is very tight for efficient initial evaporation, and the second cODT is larger to avoid inelastic three-body collisions. The final trap consists of two beams crossed in the horizontal and the vertical directions. The horizontal beam is elliptical with a horizontal waist of $65(2)\ \mu\mathrm{m}$ and a vertical waist of $35(2)\ \mu\mathrm{m}$. The vertical beam has a circular waist of $75(2)\ \mu\mathrm{m}$. These beam profiles are chosen so that the trap is oblate, with the tight axis along gravity $-\hat{z}$, to avoid trap instabilities due to the DDI~\cite{Eberlein:2005,Koch:2008}. Throughout the evaporation, the atomic dipoles are aligned along $\hat{z}$ by a constant vertical magnetic field $B_z=1.581(5)$~G. We verified for both isotopes that there are no Fano-Feshbach resonances within a range of 100~mG centered at this field~\cite{Baumann:2014ey}.  This ensures that our measurement of $a$ corresponds to the background value.

The aforementioned cODT configurations are optimized for BEC production.  We utilize the same traps in this work,  but do not evaporatively cool the gas quite to degeneracy.  In this thermal but ultracold temperature regime, the collisional dynamics of dipolar particles can be modeled by the Boltzmann equation. We apply the same evaporative cooling sequence for  $^{162}$Dy and $^{164}$Dy, and we obtain $2.7(1)\times10^5$ ($2.6(1)\times10^5$) atoms for $^{162}$Dy ($^{164}$Dy), both at 550(10)~nK and $T/T_c\approx1.7$.

To prepare for the cross-dimensional relaxation experiment, we first raise the trap depth by adiabatically ramping up the power of both beams by a factor of 2 in 0.2~s to 1.2(1)~W for the horizontal beam and 1.9(1)~W for the vertical. A tighter, deeper trap prevents evaporation after the cloud is compressed, and the new trap frequencies are [$\omega_x,\omega_y,\omega_z$] = $2\pi\times$[151(2), 70(5), 393(1)]~Hz. We then rotate the magnetic field in the $\hat{y}$-$\hat{z}$ plane to the desired angle $\beta$, where $\beta$ is the angle between the field orientation and $\hat{z}$. We ensure that the magnitude of the field remains unchanged after the rotation to within 10~mG of the initial value through rf-spectroscopy measurements of Zeeman level splittings. We repeat the experiment at three different angles $\beta=[0.0(2)^\circ,44.7(5)^\circ,90.0(2)^\circ]$, as the dipole alignment angle should affect the thermalization time scale. A valid theory that accounts for both the anisotropic DDI and the $s$-wave interaction should extract consistent scattering lengths from measurements made at different $\beta$.

The last preparatory step involves uniformly increasing the temperature of the cloud to prevent dipolar mean-field interaction energy from affecting time-of-flight (TOF) thermometry. While the contact interaction is negligible above $T_{c}$, the DDI energy  requires accurate modeling. We find that even a thermal cloud of Dy in equilibrium expands anisotropically near degeneracy, indicating that the DDI affects TOF expansion. However, we observe isotropic expansion after heating the cloud to about 1.2 $\mu$K. We parametrically heat the cloud by modulating the power of the horizontal ODT for 0.4~s at 400~Hz, nearly resonant with $\omega_{z}$. After the heating, we hold the cloud for 0.4 s to ensure thermal equilibrium, which we verify by observing isotropic expansion at 20~ms TOF. This sets the initial state of the cross-dimensional relaxation experiment with a peak atomic density of $n_0=3.7(1)\times10^{13}$~cm$^{-3}$ and $T/T_c=2.6$ for both $^{162}$Dy and $^{164}$Dy at $\beta=0^\circ$. The $^{162}$Dy densities at $\beta=45^\circ$ and $\beta=90^\circ$ are lowered by 5\% and 16\%, respectively. For $^{164}$Dy, we observe no  decrease in density at $\beta=45^\circ$ but a 27\% decrease at $\beta=90^\circ$. These losses are likely due to Fano-Feshbach resonances encountered during the magnetic field rotation~\cite{FF:footnote}. 

To drive the cloud out of equilibrium, we increase the power of the vertical ODT by a factor of 2 with a 1-ms linear ramp. The resulting trap frequencies are [$\omega_x,\omega_y,\omega_z$] = $2\pi\times$[175(3), 103(5), 393(1)]~Hz.  The induced change in the trapping potential can be considered diabatic since the ramp time is much shorter than the trap oscillation periods in the two directions, $\hat{x}$ and $\hat{y}$, that are primarily affected by the vertical beam. During the compression process, the majority of the energy is added to the most weakly confined direction $\hat{y}$, which is along the imaging beam. The trap frequency along $\hat{x}$ is also slightly increased by the vertical beam. The extra energy then redistributes among all three dimensions as the atoms undergo elastic collisions in the trap, and we record the rethermalization process by measuring $T_x$ and $T_z$ after holding the cloud for variable durations~\cite{TOF:footnote}.  To extract the $s$-wave scattering length, we compare the measured rethermalization dynamics to the numerical simulations described in the next section.  

\section{Numerical simulation}

In non-dipolar (or sufficiently weak dipolar) Bose gases, the scattering length is simply related to the rethermalization time-constant by $\tau=\alpha/\bar{n}\sigma\bar{v}_{\rm{rel}}$, where $\bar{n}$ is the averaged atom number density, $\sigma=8\pi a^2$ is the elastic collision cross section, $v_\text{rel}=\sqrt{16k_BT/\pi m}$ is the averaged relative velocity, and $\alpha$ is the mean number of collisions per particle required for rethermalization~\cite{BohnJin_PRA_89_022702_2014}. In a strongly dipolar gas, a more complicated relationship exists between the rethermalization time constant and the scattering length because $\alpha$ becomes a function of polarization.

To simulate our experiments, we solve the Boltzmann equation using the DSMC algorithm outlined in Ref.~\cite{SykesBohn_PRA_91_013625_2015}.  The goal of the computation is to simulate the nonequilibrium dynamics of Dy gas with the single free parameter $a$. 
We expect the results of the DSMC algorithm to be quantitatively accurate at temperatures well above quantum degeneracy, but below the Wigner threshold, which for bosonic Dy corresponds to the $d$-wave centrifugal barrier $\sim$250 $\mu\rm{K}$~\cite{DeMarco1999,Kotochigova2011}.

To briefly summarize, the simulation uses $N_{\rm t}$ test-particles that undergo classical time dynamics within the trapping potential, where the $i$-th test-particle has a phase-space coordinate $\left({\bf r}_i,{\bf p}_i\right)$. Interactions are included by binning test-particles into spatial volume elements before evaluating the collision probability for every pair of test particles in accordance with Boltzmann's collision integral~\cite{Bird_Molecular_Gas_Dynamics_Clarenden_Press_Oxford_UK_1994}. This computational procedure is capable of including the complete details of the dynamic trapping potentials relevant to the experiment. The crucial ingredient in our simulations is the DDI differential scattering cross section derived analytically in the first-order Born approximation in Ref.~\cite{BohnJin_PRA_89_022702_2014}. For bosons this is given by 
\begin{widetext}
 \begin{align}
\frac{d\sigma}{d\Omega}({\bf p}_{\rm rel},{\bf p}_{\rm rel}^{\prime})=\frac{a_d^2}{2}\left[-2\frac{a}{a_d}-\frac{2(\hat{{\bf p}}_{\rm rel}.\hat{\boldsymbol\varepsilon})^2+2(\hat{{\bf p}}_{\rm rel}^{\prime}.\hat{\boldsymbol\varepsilon})^2-
4(\hat{{\bf p}}_{\rm rel}.\hat{\boldsymbol\varepsilon})(\hat{{\bf p}}_{\rm rel}^{\prime}.\hat{\boldsymbol\varepsilon})(\hat{{\bf p}}_{\rm rel}.\hat{{\bf p}}_{\rm rel}^{\prime})}{1-(\hat{{\bf p}}_{\rm rel}.\hat{{\bf p}}_{\rm rel}^{\prime})^2}+\frac{4}{3}\right]^2,\label{eq:DifferentialScattering}
 \end{align}
\end{widetext}
where $\hat{\bf p}_{\rm rel}$ and $\hat{\bf p}_{\rm rel}^{\prime}$ denote the relative momenta before and after the collision~\cite{SykesBohn_PRA_91_013625_2015}. The vector $\hat{\boldsymbol\varepsilon}$ denotes the direction of the magnetic field, to which all dipoles are aligned. The scattering cross section is a function of two length scales: the $s$-wave scattering length $a$ and the dipole length scale $a_d$.

We compute a time-dependent temperature from the momentum-space widths of the phase space distribution. Away from equilibrium, this temperature can be anisotropic:
\begin{equation}
k_{\rm B}T_j = \frac{\sigma_{p_j}^2}{m},
\end{equation}
where $\sigma_{p_j} = \sqrt{\langle{ p}_j^2\rangle}$ for direction $j$, and angle brackets denote  an average over test-particles $\langle f\left({ r},{ p}\right)\rangle=\frac{1}{N_{\rm t}}\sum_{i}f\left({ r}_i,{ p}_i\right)$, i.e., $\sigma_{p_j}$ is the standard deviation of ${ p}_j$. Alternatively, one could define temperatures from the spatial distribution rather than the momentum space, but since the experiment measures TOF expansion images, we focus on the momentum space images  to enable direct comparison between theory and experiment.

\begin{figure}[t!]
\includegraphics[width=1\columnwidth]{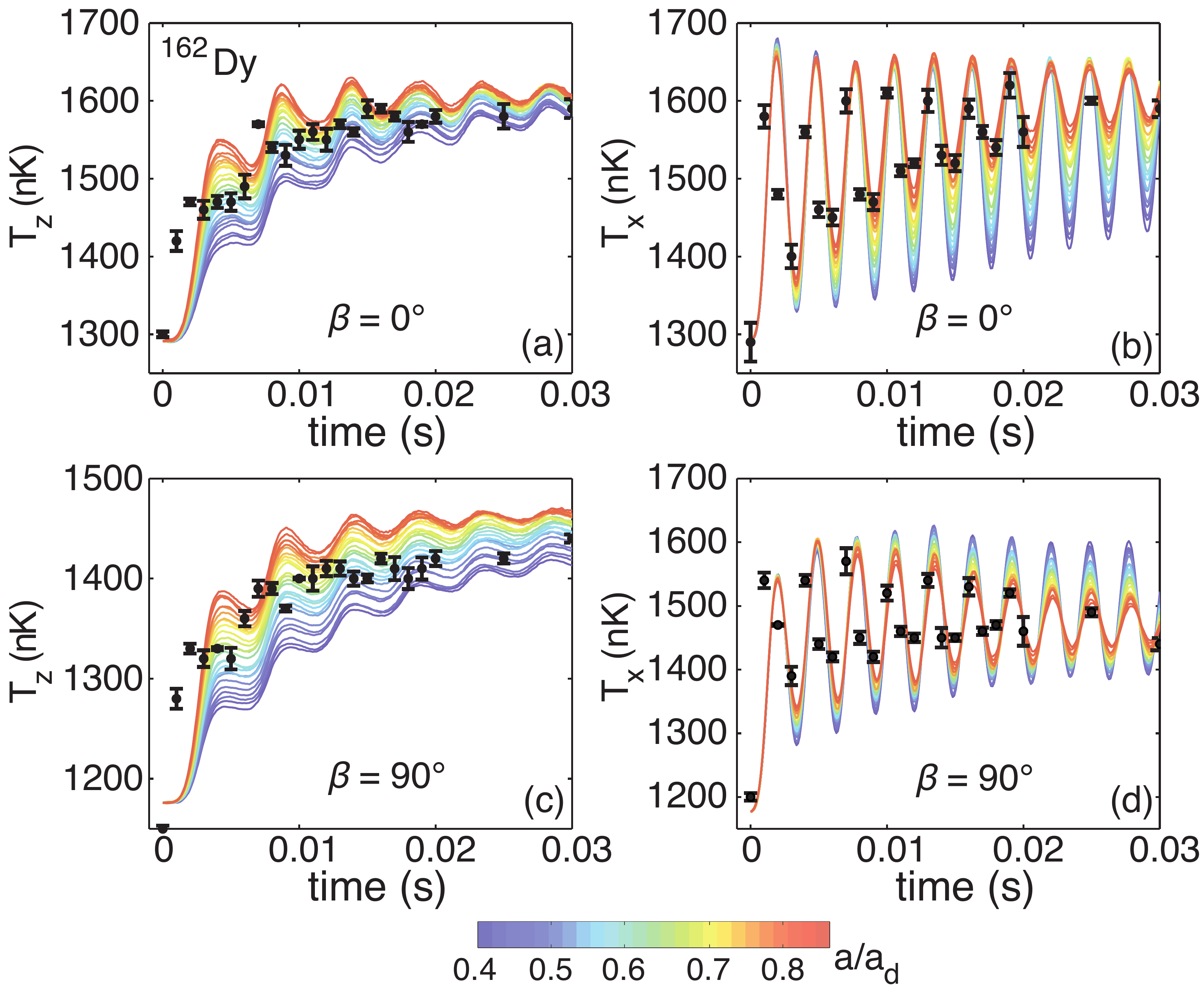}
\caption{(Color online) A qualitative comparison between the experimentally measured rethermalization of $^{162}$Dy versus results from the DSMC simulation. In (a) and (b) we show the rethermalization  dynamics for $\beta=0^\circ$, and (c) and (d) for $\beta=90^\circ$. In each plot the data points with error bars correspond to experimental measurements. In addition, there are multiple solid lines (each with a different color). These solid lines correspond to simulation results, and the color corresponds to the value of the scattering length used for simulation. The phase offset between the data and simulation is likely due to experimental uncertainty in the trap parameters.  We employ a two-step fitting method to extract estimates of the scattering length in a manner immune to these phase offsets; see Sec.~\ref{sec:fitting}. Uncertainty in these data  and in those of Fig.~\ref{fig:DataVsFittingFunction} are given as 1$\sigma$ standard errors.  Statistical fluctuations dominate systematic uncertainties in these data.}
\label{fig:DataSimulation_RawComparison}
\end{figure}

\subsection{Direct comparison between simulation and experiment}

We observe qualitative agreement between a direct comparison of experiment and simulation, some examples of which are shown in Fig.~\ref{fig:DataSimulation_RawComparison}.  The simulations use a variety of different scattering lengths to provide a visualization of the rethermalization dependence on scattering length. All curves in the simulations of Fig.~\ref{fig:DataSimulation_RawComparison} employ the same initial condition and ODT parameters. They differ only in the value of the $s$-wave scattering length.

We believe the temperature oscillations evident in Fig.~\ref{fig:DataSimulation_RawComparison} arise from collective modes excited by the diabatic trap compression.   These oscillations are unusual in  cross-dimensional rethermalization experiments, and they are due to the fact that the dysprosium gas, being highly magnetic, lies closer to the hydrodynamic collisional regime than ultracold gases of less magnetic atoms.  That is, elastic collisions occur far more frequently than in weakly dipolar gases due to the presence of both  $s$-wave  and  dipolar contributions to the elastic cross section, where the dipolar contribution is $\sigma_{DDI}=2.234a_d^2$ and $a_d\approx195 a_0$~\cite{Bohn2009}.  Indeed, our simulations show that the oscillations arise from the DDI: the oscillations vanish---and the rethermalization time increases---as the dipolar length is artificially decreased at fixed trap frequency.
The criteria for the hydrodynamic regime is $l\ll R$, where $l=1/n\sigma_\text{tot}$ is the mean-free-path, $\sigma_\text{tot}$ is the total elastic collision cross section,  and $R\sim(kT/m\omega_y^2)^{1/2}$ is the characteristic size of the gas along the weakest trap axis~\cite{pethick2002bose}.  Before compression, $l/R\approx1.5$, indicating that the collision and trapping frequencies are comparable for this highly magnetic gas.  Indeed, the oscillation frequency of $T_x$ is similar to that of $2\omega_x$, while the oscillation of $T_z$ is similar to  $2\omega_y$, the most weakly confined direction and also the direction most tightly compressed when the ODT power is abruptly increased.  

These temperature oscillations would be eliminated  by bringing the dipolar gas out of the hydrodynamic regime by reducing the trapping frequencies.  However, we cannot reduce the trap frequencies since large trap depths are required to avoid  plain evaporation of the gas after rethermalization~\cite{Grimm:1999}.  An analytic understanding of the collective excitations that  give rise to the temperature oscillations in this dipolar thermal gas are challenging and beyond the scope of the present work~\footnote{We refer the reader to  references~\cite{Yi:2002,*Goral:2002,*Ronen:2006,*Bijnen:2010,*Bismut:2010} that discuss theoretical and experimental work on collective excitations in dipolar BECs.}.

\subsection{Two-step fitting procedure}\label{sec:fitting}

We  find the frequency of the temperature oscillations to be reasonably well reproduced by the simulations.  However, the phase and amplitude seem to be highly sensitive to values of the initial and final trap frequencies as well as to the details of the ODT power ramp and are not closely replicated in the simulations.  The correspondence between simulation and data  can be improved by varying the simulated trap frequencies, total atom number,  initial temperature, and ODT ramp powers within experimental errors, but doing so for all data sets is computationally intensive. 

\begin{figure}
 \includegraphics[width=1\columnwidth]{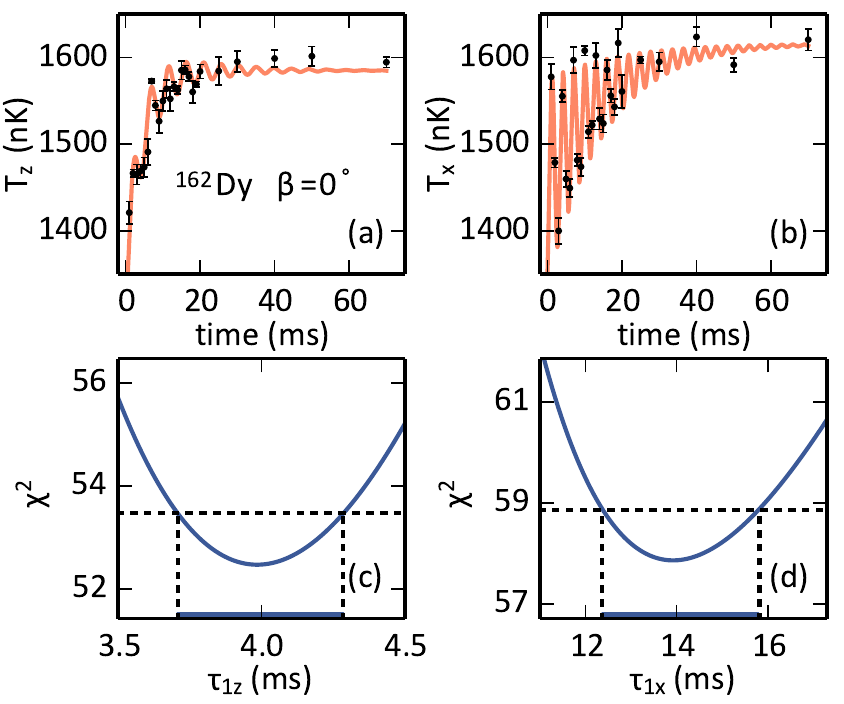}
\caption{(Color online) Fits to $^{162}$Dy data with $\beta=0^\circ$. The data points in (a) and (b) are the experimental measurements, and the solid line shows the best fit using Eq.~\eqref{eq:fitx}.  In (c) and (d) we show $\chi^2$ versus fit parameter. We vary either $\tau_{1z}$ [in (c)] or $\tau_{1x}$ [in (d)] while allowing all other parameters to be re-optimized. The blue bar along the bottom axes of (c) and (d) show the 1$\sigma$ uncertainties (where $\chi^2$ increases by 1~\cite{bevington2003data,*HughesHayes_MeasurementsAndTheirUncertainties}) in $\tau_{1z}$ and $\tau_{1x}$, respectively. Results for other $\beta$ and for $^{164}$Dy are qualitatively similar.}
\label{fig:DataVsFittingFunction}
\end{figure}

\begin{figure*}[t]
 \includegraphics[width=2\columnwidth]{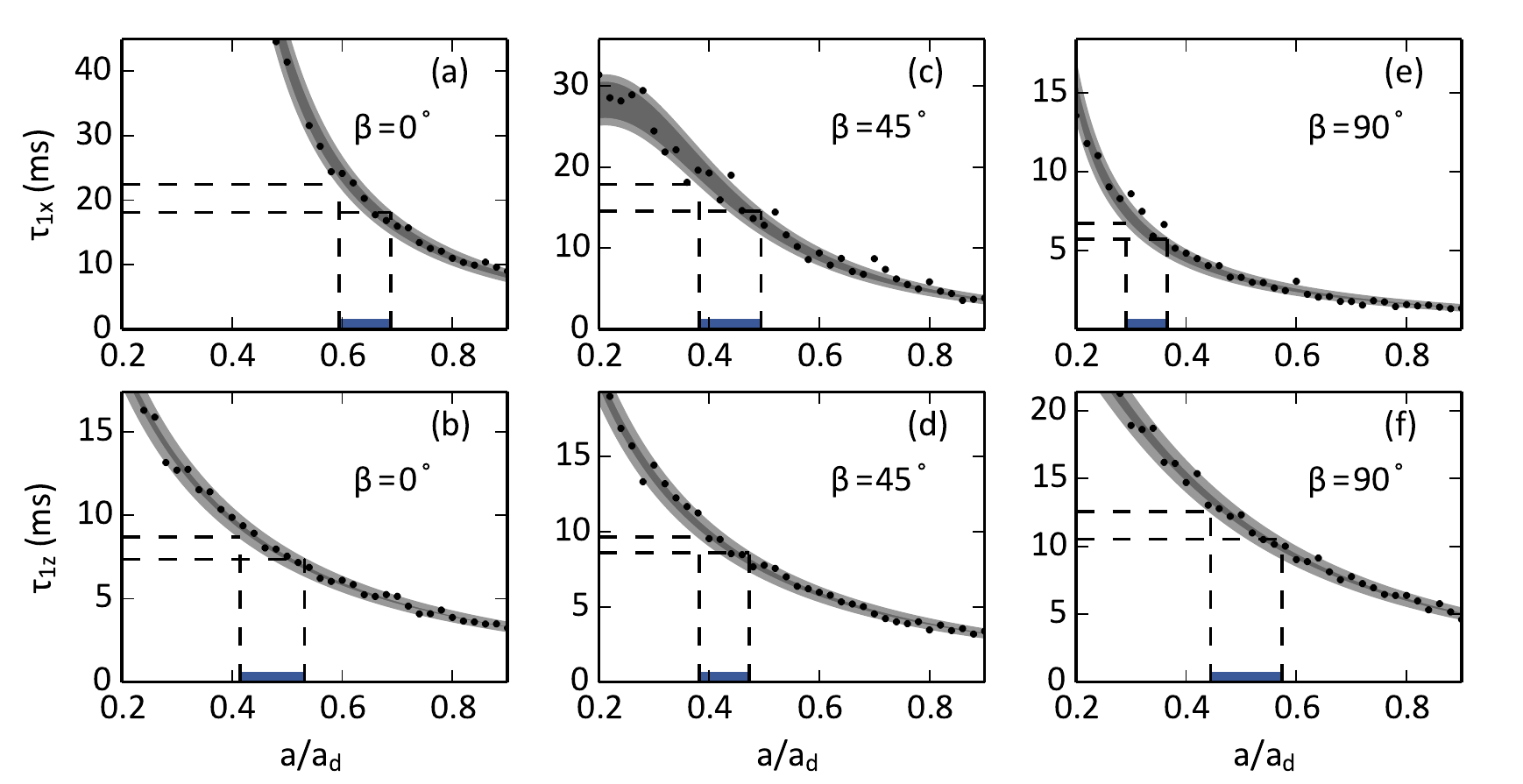}
\caption{(Color online).  Analyses of  $^{164}$Dy data. The dots show the value of $\tau_{1x,1z}$ extracted by fitting the functional form Eq.~\eqref{eq:fitx} to the simulation results. The dark grey band denotes a $1\sigma$ uncertainty on the simulated $\tau_{1x,1z}$, and the larger grey band includes experimental uncertainty. See text for  details. The horizontal dashed lines show the upper/lower bounds at $1\sigma$ uncertainty of $\tau_{1x,1z}$ found by fitting the same functional form Eq.~\eqref{eq:fitx} to the experimental data. The blue bar along the bottom axis of each figure shows the $1\sigma$ estimation of $a/a_d$, i.e., where the grey area lies between the $1\sigma$ experimental bounds. Figures (a) and (b) correspond to $\beta=0^\circ$, (c) and (d) show $\beta=45^\circ$, and (e) and (f) show $\beta=90^\circ$.  }
\label{fig:Tau1VsScatteringLength_Dy164}
\end{figure*}

\begin{figure*}[t]
 \includegraphics[width=2\columnwidth]{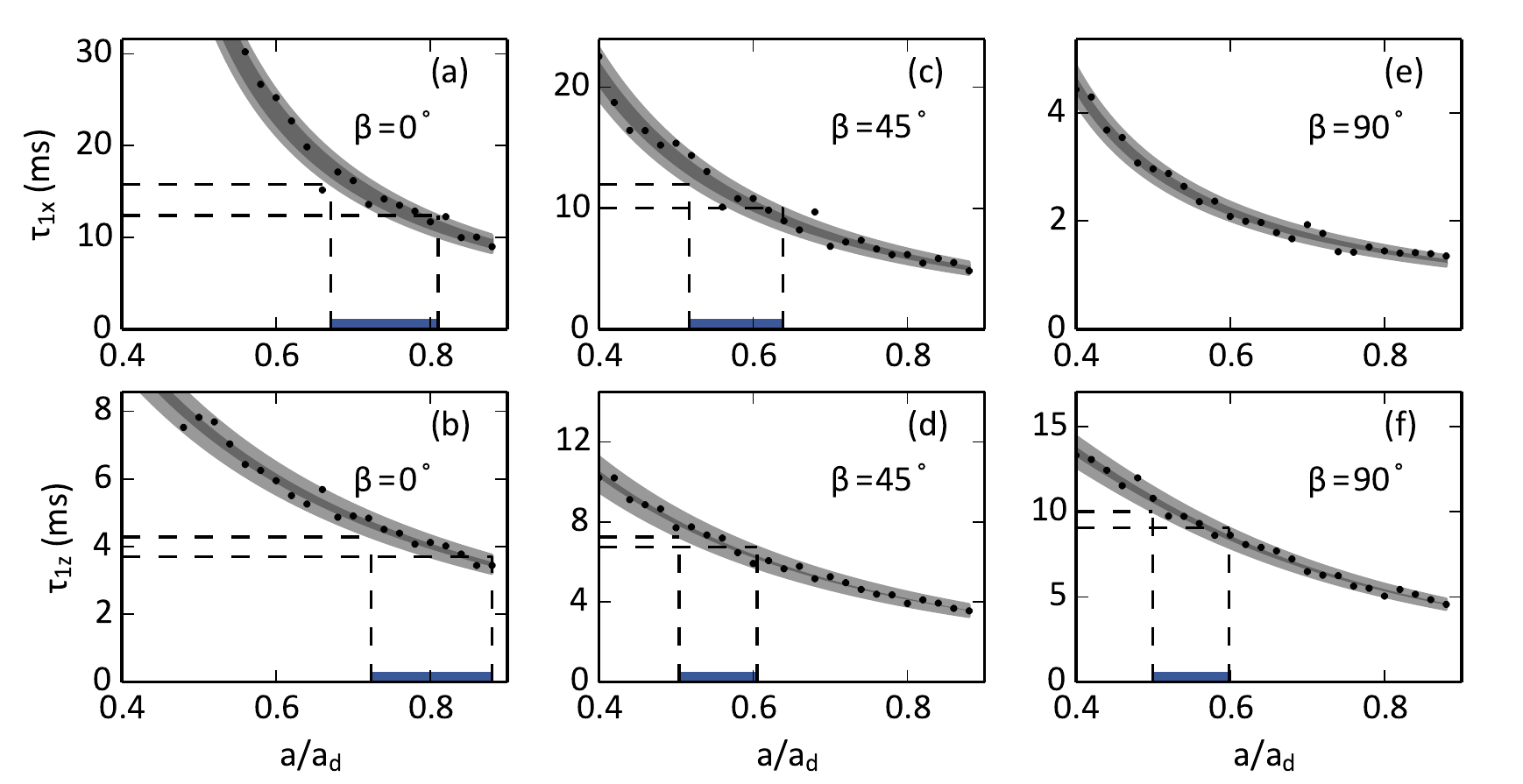}
\caption{(Color online). Analyses of $^{162}$Dy data. The dots show the value of $\tau_{1x,1z}$ extracted by fitting the functional form Eq.~\eqref{eq:fitx} to the simulation results. The dark grey band denotes a $1\sigma$ uncertainty on the simulated $\tau_{1x,1z}$, and the larger grey band includes experimental uncertainty. See text for  details. The horizontal dashed lines show the upper/lower bounds at $1\sigma$ uncertainty of $\tau_{1x,1z}$ found by fitting the same functional form Eq.~\eqref{eq:fitx} to the experimental data. The blue bar along the bottom axis of each figure shows the $1\sigma$ estimation of $a/a_d$, i.e., where the grey area lies between the $1\sigma$ experimental bounds. Figures (a) and (d) correspond to $\beta=0^\circ$, (c) and (d) show $\beta=45^\circ$, and (e) and (f) show $\beta=90^\circ$. Data for $\beta=90^\circ$ fail to  constrain $\tau_{1x}$ due to the fast thermalization time scale for $T_x$ and hence do not yield an estimate of $a/a_d$.}
\label{fig:Tau1VsScatteringLength_Dy162}
\end{figure*}

Instead, we use a two-step fitting procedure to efficiently extract estimates of $a$ from the data sets based on the observation that simulating the full equilibration evolution from first principles---oscillations of the temperature in addition to the exponential increase in temperature---is unnecessary to achieve the goal of this work. The most direct influence that $a$ has on the gas is through the scattering rate given by $\Gamma\sim \bar{n}\sigma\bar{v}_{\rm{rel}}$, which directly  contributes to the rate of equilibration in the gas.   By contrast, the temperature oscillations in the gas are more closely related to details of the trapping frequencies than to the precise value of the $s$-wave scattering length.   We may therefore extract the time constant associated with the $s$-wave cross section using a simpler model that is more robust to uncertainties in trap parameters and then use the full Boltzmann equation simulation to relate this fit parameter to the value of $a$.

We compare simulation and experiment through the function:  \begin{equation}\label{eq:fitx}
\tilde{T}_{x,z}(t)=\tilde{T}_f + (\tilde{T}_i-\tilde{T}_f) e^{-t/\tau_{1x,1z}} + \tilde{A} e^{-t/\tau_{2x,2z}} \sin\left[2\tilde{\omega} t-\tilde{\delta}\right],
\end{equation}
where this function is fit to  the experimental data (see Fig.~\ref{fig:DataVsFittingFunction}) and to the simulation results along  the $x$ and the $z$ axes separately.  The fits are restricted to times after the end of the diabatic compression ramp.  The following are free parameters: $\tilde{T}_i$ and $\tilde{T}_f$ are closely related to the initial and final temperatures, respectively; $\tau_{1x,1z}$ is the time constant for rethermalization; and $\tilde{A}$, $\tau_{2x,2z}$, $\tilde{\omega}$, and $\tilde{\delta}$  are the parameters of a damped sinusoid at the first harmonic of $\tilde{\omega}$.
 
Our fitting function reproduces both experimental and simulation results with a reduced-$\chi^2$ of order unity. We search for values of the free-parameters which generate a local minimum in the error function $= \sum_j\left[\tilde{T}_{x,z}(t_j)-T_{x,z}(t_j)\right]^2$ where $T_{x,z}(t_j)$ is derived from either the experimental measurement or the simulation. There  exist multiple local minima, but we are careful to choose the local minimum which lies nearest to the physically meaningful values of $\tilde{T}_i$, $\tilde{\omega}$, etc.

We expect, based on physical grounds, that the damping time-scales $\tau_{1x,1z}$ and $\tau_{2x,2z}$ to be the free-parameters most affected by the scattering length (through the cross section).  We now focus our attention on these two parameters. For concreteness, we continue with a description of our data analysis for the case of rethermalization along the $x$-axis; an equivalent procedure applies along the $z$-axis. Once we have found the parameters that best fit Eq.~\ref{eq:fitx} to our experimental data, we calculate a $\chi^2$ value for that fit and denote it $\chi^2_{\rm min}$. To obtain the 1$\sigma$ uncertainty on $\tau_{1x}$ and $\tau_{2x}$, we vary them while allowing all other parameters to be re-optimized until $\chi^2$ rises to $\chi^2_{\rm min}+1$~\cite{bevington2003data,*HughesHayes_MeasurementsAndTheirUncertainties}.

We find that the experimental data tightly constrain the values of $\tau_{1x}$ and $\tau_{1z}$, the parameters that characterize the overall rethermalization of the gas following the sudden squeezing of the trap. However, Fig.~\ref{fig:DataSimulation_RawComparison} shows that the experimental data are insufficient to make precise measurements of $\tau_{2x}$ and $\tau_{2z}$, which characterize the damping of the collective oscillations. Two distinct difficulties apply to the $x$-axis and $z$-axis separately: Along the $x$-axis, the 1-ms separation between the data points is comparable to the period of these oscillations, and quantitative analysis of the oscillations cannot be made due to uncertainty from under-sampling. In contrast, along the $z$-axis the oscillation frequency is well captured by the data, but the amplitude is small compared to statistical errors. Thus, we rely  on our measurements of $\tau_{1x,1z}$ for our estimates of $a$. Note that while $\tau_{2x,2z}$ do not help to constrain the value of $a$, they are consistent with the measured values of $\tau_{1x,1z}$:  We expect and observe $\tau_{2x,2z}$ to be longer than $\tau_{1x,1z}$ by approximately a factor of two as well as both time scales to be of order $1/\Gamma$~\cite{SykesBohn_PRA_91_013625_2015}. 

To assign a scattering length $a$ to each measured $\tau_{1x,1z}$, we fit the simulated rethermalization to Eq.~\ref{eq:fitx} to extract a $\tau_{1x,1z}$ for each value of $a$.  The set of these $\tau_{1x,1z}$'s are shown as dots in the panels of Figs.~\ref{fig:Tau1VsScatteringLength_Dy164} and~\ref{fig:Tau1VsScatteringLength_Dy162}. The Monte-Carlo nature of the simulation leads to an uncertainty in the predicted values for $\tau_{1x,1z}$. The resulting 1$\sigma$ uncertainties are shown as the smaller, darker grey bands in these plots. This band is found by first fitting the simulation dots to a functional form
\begin{equation}\label{fitform2}
\tau_{1x,1z}=c_1/[c_2+c_3 (a/a_d)+(a/a_d)^2],
\end{equation} which is motivated by the quadratic dependence on $a/a_d$ in the cross section; see Eq.~\ref{eq:DifferentialScattering}. We then use a bootstrap method to estimate the error on the best fit. This is done by assigning to each data point a common relative error such that the $\chi^2$ of the fit reaches the 1$\sigma$ confidence interval value of the $\chi^2$-distribution with the appropriate number of degrees of freedom~\cite{bevington2003data,*HughesHayes_MeasurementsAndTheirUncertainties}. The best-fit curve is then scaled by the estimated relative error to produce the 1$\sigma$ uncertainty represented by the dark grey band. One additional source of error on $\tau_{1x,1z}$ arises from the uncertainties in trap frequencies and atom number. This error can be determined analytically using the relation $\tau\propto1/\bar{n}$, where the mean density $\bar{n}$ contains the relevant experimental parameters. The combined 1$\sigma$ error is shown as the larger, light grey band in Figs.~\ref{fig:Tau1VsScatteringLength_Dy164} and~\ref{fig:Tau1VsScatteringLength_Dy162}. Once the relation between $\tau_{1x,1z}$ and $a/a_d$ has been established in Figs.~\ref{fig:Tau1VsScatteringLength_Dy164} and~\ref{fig:Tau1VsScatteringLength_Dy162}, one can simply project a given measured $\tau_{1x,1z}$, with its associated 1$\sigma$ uncertainty, onto the $x$-axis to obtain the best-fit $a/a_d$ value and its 1$\sigma$ uncertainty, as indicated by the horizontal and vertical dashed lines in the figures.

\section{Results}

As shown in Figs.~\ref{fig:Tau1VsScatteringLength_Dy164} and~\ref{fig:Tau1VsScatteringLength_Dy162}, the measured $\tau_{1x,1z}$'s at three different $\beta$ angles produce six independent measurements of the scattering length $a$ for each isotope, except for $^{162}$Dy at $\beta=90^\circ$.   In this case, the data fails to yield a constraint on $\tau_{1x}$.  We believe this is because we  fit to data after the 1-ms ODT ramp time, and 1~ms  is comparable to the thermalization time scale of $T_x$ at this $\beta$; see~Figs.~\ref{fig:DataSimulation_RawComparison}(d) and~\ref{fig:Tau1VsScatteringLength_Dy162}(e). The dependence of $\tau_{1x,1z}$'s on $\beta$ directly shows the anisotropic nature of the DDI: $\tau_{1x}$ decreases while $\tau_{1z}$ increases as $\beta$ is rotated from $0^\circ$ to $90^\circ$.

\begin{figure}[t]
 \includegraphics[width=1\columnwidth]{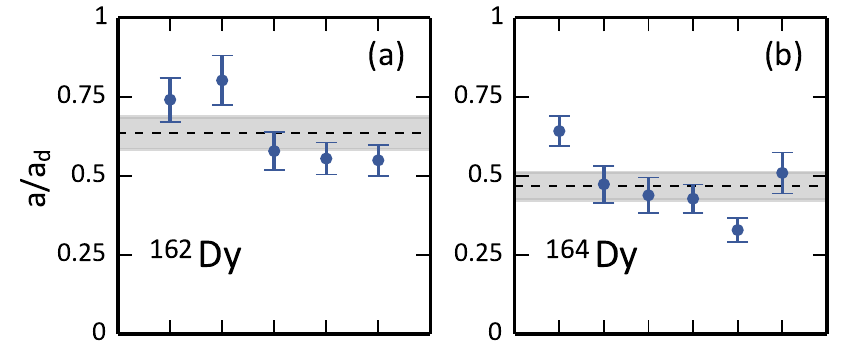}
\caption{(Color online). Summaries of the measured scattering lengths extracted from each individual experiment along with the weighted average (dashed line) and its $1\sigma$ error (grey band). Results for $^{162}$Dy and $^{164}$Dy are shown in (a) and (b), respectively. The weighted averages and $1\sigma$ standard errors are $a/a_d=0.63(5)$ for $^{162}$Dy and $a/a_d=0.47(4)$ for $^{164}$Dy.}
\label{fig:ScatteringLengths_Dy162_Dy164}
\end{figure}

The measured $a$ values are summarized in Fig.~\ref{fig:ScatteringLengths_Dy162_Dy164}. The measured values for each isotope are, in general, consistent with each other. The dashed line represents the weighted average of $a/a_d$ and the grey band represents 1$\sigma$ uncertainty calculated using the procedure described in Ref.~\cite{Paule:1982}. The weighted average values of $a/a_d$ are 0.63(5) for $^{162}$Dy and 0.47(4) for $^{164}$Dy. In absolute units, they correspond to $s$-wave scattering lengths of $a_{162}=122(10)a_0$ for $^{162}$Dy and $a_{164}=92(8)a_0$ for $^{164}$Dy. As a comparison, the mean scattering length~\cite{Gribakin:1993,Julienne:1999} is 73$a_0$, as estimated  using the value of $C_6=1890$ (a.u.) for Dy obtained via the calculations of Ref.~\cite{Kotochigova2011}.

These numbers are consistent with our previous observations regarding the different behaviors between the two isotopes. First, the larger scattering length of $^{162}$Dy could explain its  higher evaporative cooling efficiency compared to $^{164}$Dy. We were able to achieve BEC of $^{162}$Dy with an order-of-magnitude  increase in the atom number compared to $^{164}$Dy when using the same evaporation sequence~\cite{Tang:2015}. Second, the smaller $a/a_d$ value of $^{164}$Dy suggests it is more susceptible to trap instabilities due to the DDI. Previous theoretical and experimental work show that a  dipolar BEC is stable against collapse in traps with dipoles aligned along the weakest trap axis only if $a/a_d\gtrsim 2/3$~\cite{Eberlein:2005,Koch:2008,DipoleLength:footnote}. The strongly dipolar gas of $^{164}$Dy does not meet this condition, and indeed in an earlier work we found $^{164}$Dy does not form stable BEC in such a trap~\cite{Lu2011}. On the other hand, $^{162}$Dy's scattering length is close to the critical value, and we found $^{162}$Dy BECs to be stable in such traps~\cite{Tang:2015}.

We are not able to employ the above cross-dimensional relaxation procedure and analysis to measure the scattering length of the lower-abundance isotope $^{160}$Dy.   This is likely due to  either the small trap population of the gas or its small collisional cross section, or both.  The slow elastic collision rate  leads to an unreasonably long rethermalization timescale. Indeed, we observe that temperatures along $\hat{x}$ and $\hat{z}$ do not reach equilibrium before trap loss is observed, rendering Boltzmann simulations unreliable due to the violation of  equipartition.  Our previous work~\cite{Tang:2015} showed that while we could make a $^{160}$Dy BEC by tuning to a Fano-Feshbach resonance, the condensate population was only $10^3$.  No BEC could be made away from a resonance, implying that $^{160}$Dy has a background $a$ insufficient for producing stable condensates, as would typically be the case for a small and/or negative value of $a$.  Other techniques for measuring scattering lengths might prove more effective for $^{160}$Dy~\cite{pethick2002bose}.

\section{Conclusions}

We measured the rethermalization process of ultracold dipolar $^{162}$Dy and $^{164}$Dy gases driven out of equilibrium. The observed dynamics of the gases can be  described by DSMC simulations based on a Boltzmann equation that incorporates the dipolar differential scattering cross section. The agreement between experiment and theory allows us to extract the deca-heptuplet $s$-wave scattering length for both isotopes in their maximally stretched ground state. Knowledge of the scattering lengths of  $^{162}$Dy and $^{164}$Dy now allows researchers to more accurately calculate properties of these highly magnetic systems.  Such calculations are relevant to engineering and interpreting Dy-based simulations of quantum many-body physics.

We thank K.~Baumann and J.~DiSciacca for experimental assistance.   NQB and BLL acknowledge support from the Air Force Office of Scientific
Research, Grant No.~FA9550-12-1-0056, and NSF, Grant No.~1403396. YT acknowledges support from the Stanford Graduate Fellowship.   JLB and AS acknowledge support from the JILA NSF Physics Frontier
Center, Grant No.~1125844, and from the Air Force Office of Scientific
Research under the Multidisciplinary University Research
Initiative, Grant No. FA9550-1-0588.

\textit{Note added after preparation:} Using Fano-Feshbach spectroscopy, T.~Maier \textit{et al.}~\cite{Maier2015}  recently report a value of $a$ for $^{164}$Dy consistent with ours.


\begin{thebibliography}{44}
\bibitem [{FF:()}]{FF:footnote}%
  \BibitemOpen
  \href@noop {} {}\bibinfo {note} {While we use rf spectroscopy to ensure that
  the magnitude of the field is fixed to the same value $B_\beta=1.581(5)$~G at
  each of the three $\beta$'s, we allow the field magnitude to change during
  the rotation for experimental convenience.}\BibitemShut {Stop}%
\bibitem [{TOF()}]{TOF:footnote}%
  \BibitemOpen
  \href@noop {} {}\bibinfo {note} {The temperatures are measured by fitting to
  the rate of cloud expansion using TOF times 8, 10, 14, and
  16~ms.}\BibitemShut {Stop}%
\end{thebibliography}
\end{document}